\documentclass[runningheads]{llncs}
\usepackage[T1]{fontenc}
\usepackage{enumerate}
\usepackage{url}
\usepackage{graphicx}
\usepackage{booktabs}
\usepackage{tabularx}

\PassOptionsToPackage{hyphens}{url}
\usepackage[pdftex, colorlinks=true, hyperfootnotes=false, hyperindex=true, plainpages=false, pagebackref=false, pdfpagelabels=true, pdfstartview=FitH, linkcolor=blue, citecolor=blue, urlcolor=blue, bookmarks=false]{hyperref}
\usepackage{hyperref} 

\usepackage{color}

\usepackage{arydshln}

\usepackage{enumitem}
\usepackage{todonotes}
\usepackage{listings}
\usepackage{cite}
\usepackage{wrapfig}

\usepackage[subtle]{savetrees}

\makeatletter
\g@addto@macro{\UrlBreaks}{\UrlOrds}
\makeatother

\begin{document}
\title{Unlocking Sustainability Compliance: Characterizing the EU Taxonomy for Business Process Management}
\titlerunning{Characterizing the EU Taxonomy for Business Process Management}

\author{Finn Klessascheck\inst{1,3}\orcidID{0000-0001-6961-1828} \and
Stephan A. Fahrenkrog-Petersen\inst{2,3}\orcidID{0000-0002-1863-8390}\and Jan Mendling\inst{2,3}\orcidID{0000-0002-7260-524X} \and
Luise Pufahl\inst{1,3}\orcidID{0000-0002-5182-2587}}
\authorrunning{F. Klessascheck et al.}
\institute{Technical University of Munich, School of CIT, Heilbronn, Germany\\\email{firstname.lastname@tum.de} 
\and Humboldt-Universit{\"a}t zu Berlin, Berlin, Germany \\\email{firstname.lastname@hu-berlin.de} 
\and Weizenbaum Institute, Berlin, Germany 
}
\maketitle  %
\begin{abstract}
To promote sustainable business practices, and to achieve climate neutrality by 2050, the EU has developed the \emph{taxonomy of sustainable activities}, which describes when exactly business practices can be considered sustainable. While the taxonomy has only been recently established, progressively more companies will have to report how much of their revenue was created via sustainably executed business processes. To help companies prepare to assess whether their business processes comply with the constraints outlined in the taxonomy, we investigate in how far these criteria can be used for \emph{conformance checking}, that is, assessing in a data-driven manner, whether business process executions adhere to regulatory constraints. For this, we develop a few-shot learning pipeline to characterize the constraints of the taxonomy with the help of an LLM as to the \emph{process dimensions} they relate to. We find that many constraints of the taxonomy are useable for conformance checking, particularly in the sectors of energy, manufacturing, and transport.
This will aid companies in preparing to monitor regulatory compliance with the taxonomy automatically, by characterizing what kind of information they need to extract, and by providing a better understanding of sectors where such an assessment is feasible and where it is not.

\end{abstract}

\section{Introduction}

In light of the issue of climate change and unsustainable human activity~\cite{unenvironmentGlobalEnvironmentOutlook2019}, it is important to promote \emph{sustainable} business practices, i.e., conducting business in a way that can meet the needs of the present generations without endangering those of the future~\cite{brundtlandOurCommonFuture1987}. %
This need has also been identified by governing bodies, such as the \emph{European Union} (EU).
As a consequence, the EU has defined the \emph{taxonomy for sustainable activities}~\cite{europeancommissionEUTaxonomySustainable,europeancommissionRegulation202085}, subsequently referred to by us as \emph{taxonomy}. The taxonomy aims to create clear indicators of when business activities are contributing towards sustainability and when not, and to create financial incentives for proven sustainable business practices and investments into them~\cite{mcclellanEUTaxonomyReporting2023,schutzeEUSustainableFinance2024}.
For various \emph{business practices},\footnote{Referred to as \emph{economic activities} in the taxonomy; to avoid confusion with the notion of business process activities we use the term \emph{business practice}.} the taxonomy defines criteria along which a \emph{substantial contribution} towards \emph{sustainability goals} can be verified, and criteria which must not be violated. %
Increasingly, companies \emph{will} face having to assess their business processes for compliance with the taxonomy~\cite{mcclellanEUTaxonomyReporting2023}.
However, assessing whether a business practice does or does not meet relevant criteria is, so far, a manual process: Some companies offer manual or semi-automatic questionnaire-based assessments; a \emph{taxonomy calculator} provided by the EU relies exclusively on manual input in the form of an Excel sheet.\footnote{See \url{https://viridad.eu}, \url{https://www.briink.com/solutions/esg-questionnaire-assistant} and \url{https://ec.europa.eu/sustainable-finance-taxonomy/wizard} [Accessed: 23/05/2024]}

To overcome the challenges of manually assessing whether a business activity meets the criteria of the taxonomy, it appears feasible to \emph{check in a data-driven manner} whether the execution of a business practice complies with relevant taxonomy criteria or not.
Since the definition of business practice \cite{europeancommissionNACERev2008} is closely related to that of business processes \cite{weskeBusinessProcessManagement2012}, we interpret business practices as “categories” of business processes, which allows us to investigate the taxonomy and its criteria from a \emph{business process management} (BPM) standpoint.

\begin{figure}[htbp]
    \centering
    \includegraphics[width=0.9\textwidth]{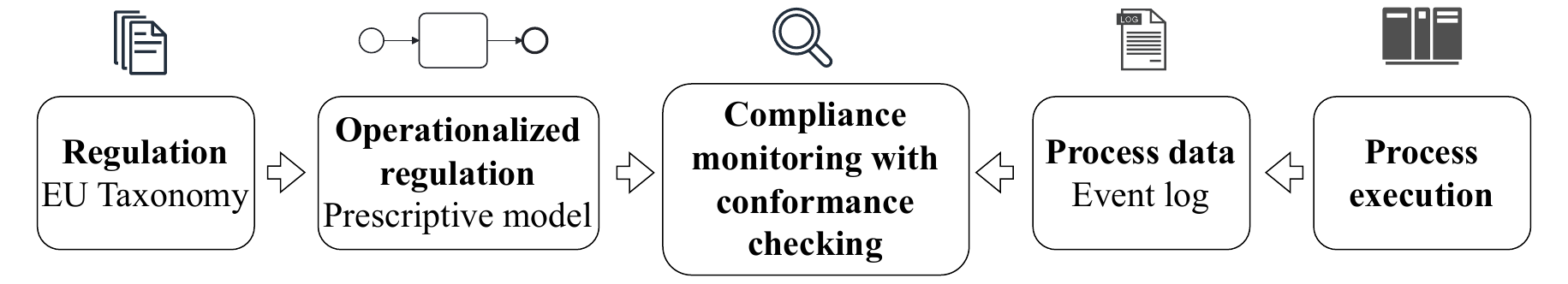}
    \caption{Conceptual overview of compliance monitoring with conformance checking~\cite{carmonaConformanceCheckingRelating2018,klessascheckReviewingConformance2024}}
    \label{fig:cc-overview}
\end{figure}

\emph{Conformance checking}, which is a technique of the \emph{process mining} and BPM fields, aims to compare recorded business process executions in the form of an event log with a formal representation of the to-be process behavior, so that either, the process execution can be improved to more closely resemble the formal representation, or vice versa~\cite{carmonaConformanceCheckingRelating2018}. Automatic compliance monitoring can use conformance checking techniques with the goal of assessing whether a business process complies with regulatory constraints --- such as those described in the taxonomy --- during its execution, based on recorded event data~\cite{groefsemaUseConformanceCompliance2022,klessascheckReviewingConformance2024}. An overview of this is provided in Fig.~\ref{fig:cc-overview}. For applying conformance checking with the aim of monitoring compliance to the regulatory constraints described in the taxonomy, this taxonomy first needs to be \emph{operationalized}, that is, translated into the form of a \emph{prescriptive model}. Further, the prescriptive model and the event data used for conformance checking need to align w.r.t. what information they contain. For this, companies need to be aware of what data they need to capture during the execution of their business processes. Therefore, we see a need to better understand what data-capturing requirements the taxonomy imposes on business processes, and in how far the constraints contained therein are even applicable for conformance checking.

\vspace{-2em}

Related work has focussed on extracting constraints from text directly into process models (e.g.,~\cite{winterDerivingCombiningMixed2019,vanderaaExtractingDeclarativeProcess2019,mustrophVerifyingResourceCompliance2023}) and on characterizing textual constraints based on i.a. their relation to other constraints, process models, or their relevance to a given business process (e.g.,~\cite{winterDetectingConstraintsTheir2018,winterAssessingComplianceBusiness2020,saiWhichLegalRequirements2024}). Our investigation, however, aims at an earlier stage of conformance checking that does not require operationalization into a concrete prescriptive model or a concrete business process against which conformance is to be checked. Rather, we aim to understand what kind of requirements the taxonomy imposes on event log data so that it can be captured and prepared for regulatory compliance checking with conformance checking w.r.t. the taxonomy. For this, we first need to understand what kinds of constraints the taxonomy embodies (i.e., which characteristics a prescriptive model would to make prescriptions towards, such as certain activities that need to be executed in a specific order, certain thresholds that activities must not exceed, etc.), and whether all of them can be related to a process view, since so far, it is unclear to what extent the taxonomy can even be operationalized for conformance checking. 
Therefore, this work aims to address the following \emph{research questions} (RQ):

\begin{enumerate}[label=\textbf{RQ\arabic*:},wide=0pt, leftmargin=*] %
    \label{RQs}
    \item \label{RQ1} How can the EU taxonomy be operationalized with regard to business processes?
    \item \label{RQ2} Which constraints of the EU taxonomy can be used for automatically assessing whether a business process fulfills its respective sustainability criteria with conformance checking techniques?
\end{enumerate}

\noindent
Since the taxonomy contains constraints for around 80 types of business practices~\cite{schutzeEUSustainableFinance2024,malecki2021eu} -- which makes a manual characterization of the entire taxonomy infeasible -- this work uses a \emph{few-shot machine learning approach} to gain insights into what constraints the taxonomy consists of, and how they might be operationalized in practice. In doing so, we also explore the potential of novel approaches based on \emph{large language models} (LLMs) for operationalizing regulations in the area of conformance checking.

The remainder of the article is organized as follows: Section~\ref{sec:background} provides background on the taxonomy, regulatory compliance monitoring with conformance checking, and few-shot learning approaches. Section~\ref{sec:related} provides related work on approaches for extracting rules from text for compliance monitoring. In Sec.~\ref{sec:methods}, we present the research approach of this paper. Section~\ref{sec:experimental} provides the results thereof, characterizing the constraints of the taxonomy and its potential for conformance checking uses. 
We further discuss their implications for practice in Sec.~\ref{sec:discussion}. 
Finally, Sec.~\ref{sec:conclusion-futur-work} provides future work and concludes the article.

\section{Background}
\label{sec:background}
Next, we will outline the general idea of the EU taxonomy, and we will give an overview of automatic compliance monitoring approaches within the business process literature.

\subsection{EU Taxonomy for Sustainable Activities}

In order to create incentives for investments in sustainable technologies and to provide a transparent classification system for when business practices are sustainable, the EU has established the \emph{taxonomy for sustainable activities}~\cite{schutzeEUSustainableFinance2024,alessiEUSustainabilityTaxonomy2019,europeancommissionEUTaxonomySustainable}. 
The ultimate objective behind the taxonomy is to support the EU in transiting to climate neutrality by 2050~\cite{schutzeEUSustainableFinance2024,mcclellanEUTaxonomyReporting2023}.

\begin{figure}[htbp]
    \centering
    \includegraphics[width=0.8\textwidth]{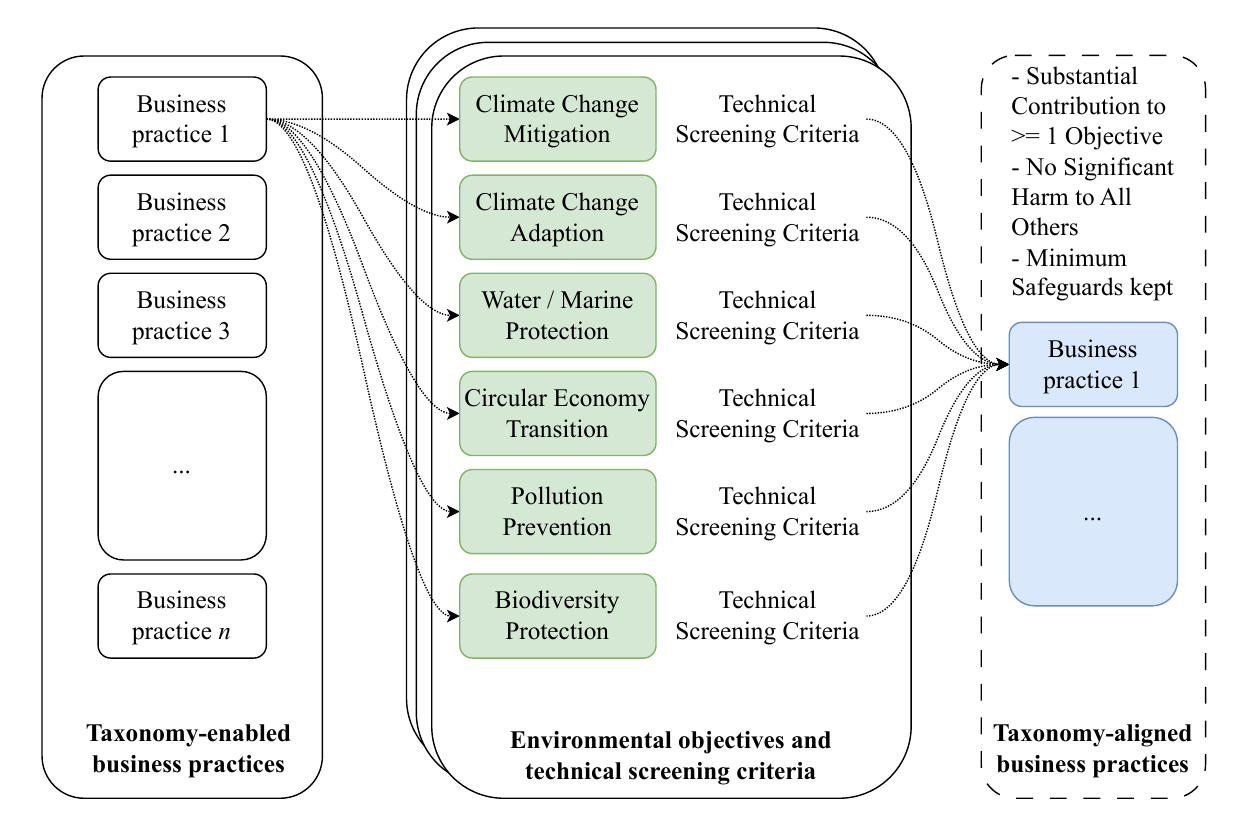}
    \caption{Schematic overview of the EU taxonomy for sustainable activities and its concepts, derived from~\cite{europeancommissionEUTaxonomySustainable,europeancommissionUserGuideNavigate2023}}
    \label{fig:taxonomy-overview}
    \vspace{-1em}
\end{figure}

In essence, the taxonomy defines six \emph{environmental objectives} with which sustainable business practices are identified: 1) mitigating climate change; 2) adapting to climate change; 3) sustainably using and protecting water and marine resources; 4) transitioning to a circular economy; 5) preventing and controlling pollution; and 6) protecting and restoring biodiversity and the ecosystem~\cite{conea2022eu,malecki2021eu}.
Not all possible business practices of all industries are covered by the taxonomy, but only those which are deemed to be able to make a \emph{substantial contribution} towards climate neutrality, or are needed for other sectors to make a substantial contribution~\cite{schutzeEUSustainableFinance2024}.
If a business practice is part of the taxonomy, it is called taxonomy-\emph{enabled}, and can \emph{potentially} make a contribution to one of the six environmental objectives. If a taxonomy-enabled business practice: 1) indeed contributes to one of the environmental objectives; 2) causes no \emph{significant harm} (DNSH) to any of the six objectives; 3) meets \emph{minimum safeguards} (such as the UN Guiding Principles on Business and Human Rights); 4) adheres to \emph{technical screening criteria}, it is, in fact, \emph{sustainable} according to the taxonomy~\cite{alessiEUSustainabilityTaxonomy2019,conea2022eu,malecki2021eu}. Business practices that meet all of these criteria are also called \emph{taxonomy-aligned}.
In short, to be taxonomy-aligned, a taxonomy-enabled business practice \emph{must} contribute to at least one of the six environmental objectives, \emph{must not} cause significant harm to any of the others, and \emph{must} meet minimum safeguards~\cite{alessiEUSustainabilityTaxonomy2019,conea2022eu,malecki2021eu}. Notably, the minimum safeguards are not directly defined \emph{in} the taxonomy, but rather, are references to taxonomy-external guidelines and regulations. Concretely, organizations need to ensure that they follow the \emph{OECD Guidelines for Multinational Enterprises}, \emph{UN Guiding Principles on Business and Human Rights}, the \emph{Declaration of the International Labour Organisation on Fundamental Principles and Rights at Work} as well as the \emph{International Bill of Human Rights} \cite[Article 18]{europeancommissionRegulation202085}. Since the focus of our work is on compliance of business processes and not of organizations or supply chains, and minimum safeguards are often not assessed on the level of business processes \cite{mcclellanEUTaxonomyReporting2023}, we forgo including these external constraints in our investigation. Figure~\ref{fig:taxonomy-overview} provides a schematic overview of the taxonomy and its concepts.

For assessing the taxonomy alignment of a taxonomy-enabled business practice, the taxonomy describes \emph{technical screening criteria}, which describe under which circumstances an activity either makes a contribution or causes significant harm to an environmental objective~\cite{schutzeEUSustainableFinance2024,malecki2021eu}. It should be noted that a business practice is not necessarily able to make substantial contributions to more than one economic objective, and hence may have only one set of technical screening criteria for one substantial contribution.

For an illustration of how the taxonomy documents technical screening criteria for environmental objectives, we refer to the EU's \emph{taxonomy compass} and the corresponding Excel file.\footnote{See \url{https://ec.europa.eu/sustainable-finance-taxonomy/taxonomy-compass/the-compass} and \url{https://ec.europa.eu/sustainable-finance-taxonomy/assets/documents/taxonomy.xlsx} [Accessed: 18/06/2024]} Generally, for each business practice, a set of possible environmental objectives to which a substantial contribution can be made is provided --- for each objective, criteria for a substantial contribution, as well as DNSH, are documented.

By determining how many of their business processes align with the taxonomy, companies can report how much of their business output is generated via sustainable business practices~\cite{mcclellanEUTaxonomyReporting2023,europeancommissionUserGuideNavigate2023}.
For this, companies need to know exactly which of their business practices are not just taxonomy-enabled, but also taxonomy-aligned --- and in the future, these taxonomy disclosures will be subject to mandatory audits~\cite{mcclellanEUTaxonomyReporting2023}.
Notably, from 2025 onwards, companies that meet certain characteristics (i.e., more than 250 employees, more than EUR 25M balance sheet total, or more than EUR 50M net turnover) will be obligated to do so, with the threshold for having to report decreasing in subsequent years~\cite{mcclellanEUTaxonomyReporting2023}.
Small and medium-size enterprises will also be subject to a disclosure obligation~\cite{mcclellanEUTaxonomyReporting2023}.
Therefore, it appears prudent to investigate how automatic business process compliance monitoring can help companies in assessing the taxonomy-alignment of their business processes.

\subsection{Automatic Business Process Compliance Monitoring}
The automatic monitoring of business process compliance~\cite{lopezBusinessProcessCompliance2020} allows organizations  to ensure their business practices comply with regulations~\cite{hashmiAreWeDone2018}. Through an analysis of process execution data, it is possible to check at runtime if a business process complies with specified rules~\cite{groefsemaUseConformanceCompliance2022,klessascheckReviewingConformance2024}. 
Figure~\ref{fig:cc-overview} provides a conceptual overview of automatic compliance monitoring with conformance checking, with the EU taxonomy being one example of a regulation that can be operationalized. In order for an organization to adopt these techniques, the following steps need to be taken~\cite{carmonaConformanceCheckingRelating2018,klessascheckReviewingConformance2024}:

\begin{enumerate}
    \item The relevant piece of regulation needs to be operationalized into a \emph{prescriptive model}. This model embodies constraints towards one or more \emph{process dimensions}~\cite{russellWorkflowPatternsDefinitive2015}, which are \emph{control-flow}, \emph{time}, \emph{resources}, and \emph{data}~\cite{carmonaConformanceCheckingRelating2018,klessascheckReviewingConformance2024}, of the process under investigation.
    \item An \emph{event log} needs to be extracted from recorded process executions, describing the actual process behavior~\cite{remy2020event}.
    \item The prescriptive model and event log are used by a conformance checking \emph{algorithm}, which checks whether or where the process deviated from the prescriptive model. This can serve as a starting point for further diagnosing or explaining deviations, and subsequently, remedying unwanted deviations~\cite{klessascheckReviewingConformance2024,carmonaConformanceCheckingRelating2018,rehseProcessMiningMeets2022}.
\end{enumerate}

\noindent
It should be noted, however, that the prescriptive model and event log need to align in their respective process dimensions: If, for example, the prescriptive model imposes constraint on the data dimension of the investigated process, but the event log does not contain this information, the conformance check \emph{cannot} yield the desired insights~\cite{carmonaConformanceCheckingRelating2018}.
Hence, organizations need to know, potentially in advance, which process dimensions are, in fact, \emph{relevant} for the conformance check. Based on this, they can appropriately capture the relevant execution data and extract it into the subsequent event log.

\section{Related Work}
\label{sec:related}

In this paper, we investigate the properties of the EU taxonomy for sustainable business practices and its potential for being used for compliance monitoring with conformance checking, by extracting insights from the taxonomy's regulatory texts and the compliance constraints described therein. In that respect, our work relates to other contributions that also deal with rule extraction for conformance checking/process mining purposes, and contributions that use machine learning techniques to do so.

\paragraph{Constraint Extraction from Text.}

For extracting compliance constraints from regulatory documents, Dragoni et al.~\cite{dragoniCombiningNaturalLanguage2018} propose a pipeline that combines multiple \emph{natural language processing} (NLP) approaches. With their proposed pipeline, rules (in this case in the form of \emph{obligations}, \emph{permissions}, and \emph{prohibitions}, see Hashmi et al.~\cite{hashmiNormativeRequirementsRegulatory2016}) can be extracted into formal representations for a given regulatory text. Using semantic annotations, a process model can then be checked for whether it complies with the formal representations -- this check is further described by Governatori et al. ~\cite{governatoriSemanticBusinessProcess2016}. %
Moreover, Winter and Rinderle-Ma~\cite{winterDerivingCombiningMixed2019} describe an approach to generate process model fragments from regulatory documents by extracting constraints and their relation.
Similarly, van der Aa et al.~\cite{vanderaaExtractingDeclarativeProcess2019} describe an automatic approach for extracting declarative process models from natural language text that describes a process.
Barrientos et al.~\cite{barrientosVerificationQuantitativeTemporal2023} design an approach for extracting temporal constraints from natural language texts and determining violations thereof in an event log.
Focussing on resource compliance, Mustroph et al.~\cite{mustrophVerifyingResourceCompliance2023} extract compliance requirements from natural language with GPT-4, which are then matched to and verified against an event log.
Further, Mustroph et al.~\cite{mustrophSocialNetworkMining2024} describe how generative AI, or more specifically, GPT-4, can be used to pre-process resource-related regulations into social network graphs. Based on these, they are able to detect compliance violations of process executions.

\paragraph{Characterization of Constraints.}

In terms of characterizing constraints present in regulatory documents, Winter et al.~\cite{winterCharacterizingRegulatoryDocuments2017} provide a technique for characterizing regulations based on text mining and clustering algorithms that can derive significant sentences for a regulatory document, which can be manually translated into e.g. process models.
Similarly, Winter and Rinderle-Ma~\cite{winterDetectingConstraintsTheir2018} design an approach to group constraints from textual documents and detect relations between them, e.g. based on similarity.
Moreover, Winter et al.~\cite{winterAssessingComplianceBusiness2020} provides a method for matching parts of regulatory documents with process models, which then allows a compliance assessment of the matched regulatory constraints and the process model.
Aiming to facilitate a better understanding of regulatory documents, Sai et al.~\cite{saiIdentificationVisualizationLegal2023} propose an approach to parse legal definitions and relations between terms from regulatory documents into a knowledge graph, so that regulatory documents can be better understood and analyzed.
Further, in order to compare regulatory documents with their translation into company-internal requirements, Sai et al.~\cite{saiDetectingDeviationsExternal2023} provide an NLP-based which can detect deviations between the two documents, and helps to detect root causes of textual deviations.
Finally, Sai et al.~\cite{saiWhichLegalRequirements2024} investigate an automated approach for identifying passages of regulatory texts that are relevant for a business process based on the processes textual description. They find that expert judgement cannot be replaced by generative AI, but see the potential of AI uses for taking into account vaster amounts of context, which the authors deem to be advantageous in more complex settings.

\bigskip
\noindent
In contrast to these contributions, which primarily focus on extracting constraints directly or characterizing them w.r.t a process model, event log or other texts, we exclusively focus on characterizing the constraints for the process dimensions they constrain. This would facilitate data extraction of recorded process execution and translation of the taxonomy into concrete constraints for subsequent analyses. However, with our categorization as a starting point, relevant extraction and matching approaches can be chosen, and the relevant data can be stored during process execution.

Notably, there is currently no automated approach that helps companies to understand the requirements posed on business process executions by regulatory texts w.r.t. the process dimensions so that the log and the subsequent prescriptive model pertain to the relevant perspectives and contain relevant information for compliance monitoring with conformance checking. %
In particular, the taxonomy has not yet been considered in this light --- however, concepts from existing approaches can inform the design of new mechanisms, such as ours. 

\section{Mapping Sustainability Regulation to Conformance Constraints}
\label{sec:methods}

In order to operationalize the taxonomy for business processes, we need to identify process constraints within the taxonomy. This allows us to transform the abstract rules for business practices into specific problems that can be addressed using automatic conformance monitoring. 

We assume that it would be best to use an industry expert with conformance checking knowledge to classify each rule of the taxonomy. However, since such experts are rare and the taxonomy covers many varying industries, an alternative solution becomes necessary. Because of this, we decide to use LLMs, since they are trained on a wide variety of texts from different domains. Furthermore, our task can be framed as a text classification problem, and it was shown that LLMs can be used as \emph{Few-Shot-Classifiers}~\cite{brown2020language}. Meaning that an LLM, which is trained for one task such as next text token prediction, can be with sufficient instructions utilized to perform another machine learning task, in our case text classification of the taxonomy. In particular, previous work has shown that LLMs can be successfully be used for process analysis tasks~\cite{grohsLargeLanguageModels2024}. 

In order to characterize the taxonomy, which consists of several regulatory texts for each business practice and environmental objective it considers, we need to apply an LLM to each regulatory text and extract relevant information about the constraints. For this, we developed a pipeline, which we outline in the subsequent \autoref{sec:approach_overview}, drawing on the capability of LLMs to be applied in this manner.

\subsection{Overview}
\label{sec:approach_overview}

\begin{figure}[htbp]
    \centering
    \includegraphics[width=0.8\textwidth]{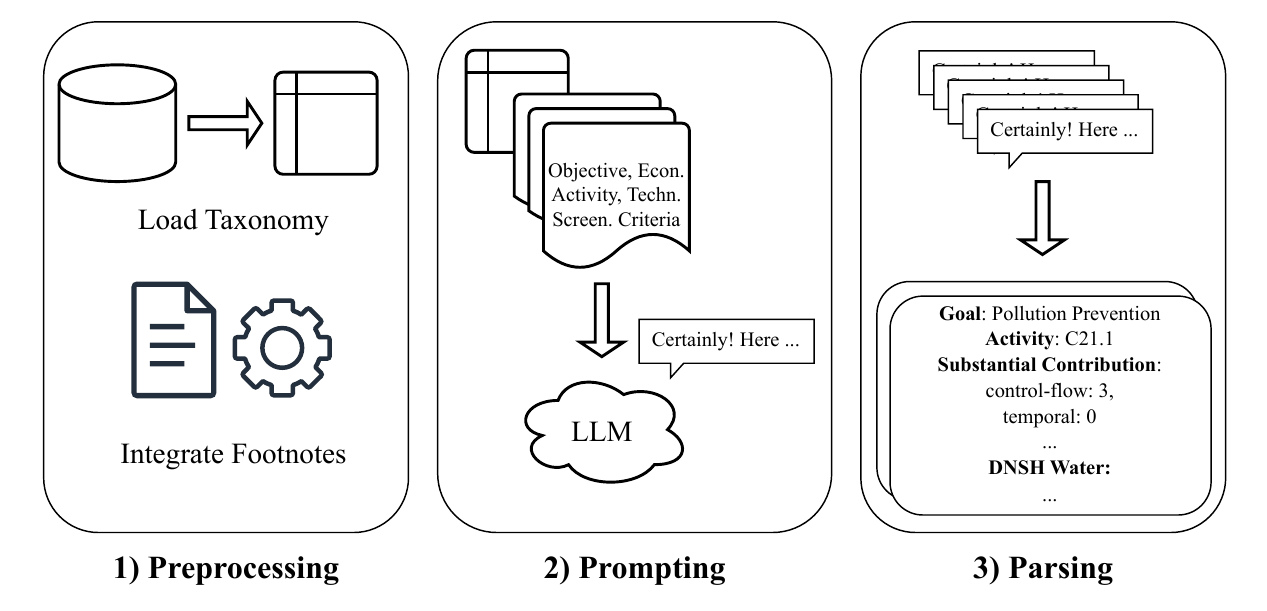}
    \caption{Schematic overview of the taxonomy constraint characterization pipeline}
    \label{fig:pipeline-overview}
\end{figure}

The overall approach for characterizing the constraints, or technical screening criteria, of the taxonomy for the process dimensions they pertain to, is illustrated in Fig.~\ref{fig:pipeline-overview}. The pipeline we describe consists of three stages: First, the taxonomy is \emph{preprocessed} (see \autoref{sec:results-preprocessing}). Second, for each set of technical screening criteria of each business practice and environmental objective, an LLM is \emph{prompted} to identify the types of constraints contained therein (see \autoref{sec:results-prompting}). Third, the LLM's output is \emph{parsed}, and the numbers and types of constraint per business practice and environmental objective is extracted (see \autoref{sec:results-parsing}). Finally, the resulting data is collected and can be \emph{analyzed}.

\subsection{Preprocessing}
\label{sec:results-preprocessing}
For the preprocessing step, we begin to read the entire taxonomy, which is available in the form of an Excel file.\footnote{\url{https://ec.europa.eu/sustainable-finance-taxonomy/assets/documents/taxonomy.xlsx} [Accessed: 18/06/2024]} For each environmental objective, the taxonomy file contains a sheet, in which each environmental activity as well as 1) the corresponding technical screening criteria for a substantial contribution to the objective, and 2) the technical screening criteria for DNSH to all other objectives are listed. Further, footnotes pertaining to each activity are contained in a column in each sheet. In order to make all relevant information available to the subsequent prompting step, the footnotes are appended to each technical screening criteria block of an activity if they are referenced therein.

\subsection{Prompting}
\label{sec:results-prompting}
Subsequently, the approach iterates through the taxonomy in the following manner:
For each of the six climate objectives, and for each business practice that can make a substantial contribution to that objective, the taxonomy contains: 1) a text that describes the technical screening criteria with which a substantial contribution can be determined, and 2) up to five texts that describes the technical screening criteria with which significant harm to the other objectives can be determined. For each of these texts, the approach prepares a prompt to an LLM, with which the types of constraints and numbers thereof present in the respective technical screening criteria can be determined. The prompt includes a small task description and a brief explanation of the process constraint types, as well as the text passage that is currently being characterized.

We differentiate the process constraints along two aspects: 
(i) We classify what \textbf{process dimension}~\cite{russellWorkflowPatternsDefinitive2015,carmonaConformanceCheckingRelating2018} is targeted by a constraint. We categorize constraints into one of the following dimensions: \emph{control-flow}, \emph{temporal}, \emph{resource}, and \emph{data}. Alternatively, a constraint can be \emph{irrelevant} from a process perspective; 
(ii) Furthermore, we distinguish the \textbf{granularity} of a constraint: meaning a constraint can either be targeted towards an aspect \emph{within} an activity or \emph{between} activities. 

An example of a \emph{temporal} constraint \emph{between} activities could be the requirement to perform one specific activity within one week after another activity was performed, while an example of a \emph{resource} constraint \emph{within} an activity is an activity that needs to be performed by a person with a specific certification.
Constraints classified as pertaining to the \emph{control-flow} \emph{within} an activity can be understood as \emph{activity existence} constraints.

\medskip
\noindent
The prompt to retrieve this information is structured as follows: 

\begin{itemize}
    \item Briefly, the overall objective is described;
    \item The different types of process constraints are described, as well as the difference in granularity;
    \item Examples consisting of excerpts of the taxonomy and their potential classification are provided;
    \item The exact task of characterizing a section of the taxonomy is described;
    \item Requirements regarding the output are described;
     \item A placeholder is provided where, during pipeline execution, a description of the business practice can be inserted; and
     \item A placeholder for the text passage of the taxonomy for which the characterization needs to be done is provided, which will be filled during pipeline execution.
\end{itemize}
The entire prompt template is available online.\footnote{\url{https://github.com/fyndalf/unlocking-sustainability-compliance-replication-package}} As a result, we  receive the following information as a response from the LLM, giving us insights into the process constraints covered by a particular set of criteria:

\begin{itemize}
    \item \# of \textbf{activity existence} constraints of process activities
    \item \# of \textbf{control-flow constraints} \emph{between} process activities
    \item \# of \textbf{temporal constraints} \emph{within} process activities
    \item \# of \textbf{temporal constraints} \emph{between} process activities
    \item \# of \textbf{resource constraints} \emph{within} process activities
    \item \# of \textbf{resource constraints} \emph{between} process activities
    \item \# of \textbf{data constraints} \emph{within} process activities
    \item \# of \textbf{data constraints} \emph{between} process activities
    \item \# of \textbf{process-irrelevant} constraints
\end{itemize}

\subsection{Parsing}
\label{sec:results-parsing}
Each response of the LLM to a prompt of one text (that describes a set of technical screening criteria) is parsed individually. The prompt instructs the LLM to return the results in a particular \emph{JSON}-like notation. This is depicted in Listing~\ref{lst:format}. 

\begin{lstlisting}[caption={Excerpt of the prompt template, describing the required response structure},label=lst:format]
    {'control-flow': {
        'within_activities': [no. of activity existence constraints],
        'between_activities': [no. of control-flow constraints between activities]},
    'temporal': {
        'within_activities': [no.  of temporal constraints within activities],
        'between_activities': [no. of temporal constraints between activities]},
    'resource': {
        'within_activities': [no. of resource constraints within activities],
        'between_activities': [no. of resource constraints between activities]},
    'data':{
        'within_activities': [no. of data constraints within activities],
        'between_activities': [no. of data constraints between activities]},
    'irrelevant': [no. of process-irrelevant constraints]}
\end{lstlisting}
\noindent
In addition to some further processing steps (such as replacing comment-like symbols), this structure and the number of the respective constraints are extracted from the response. Additionally, we store the entire response text, as it often contains the LLMs ``explanation'' for the constraints that were identified. As a result, we know the number of constraints and types of one set of technical screening criteria for one business practice and one environmental goal. The parsing step is repeated for all further sets of technical screening criteria and prompt responses of each activity. Ultimately, we end up with six datasets, one for each environmental objective. Each dataset contains information on the number of constraints imposed on each business practice by the substantial contribution and DNSH criteria, which, once collated, subsequently allows further analyses.

\section{Experimental Validation}
\label{sec:experimental}
In this section, we use our approach to validate the extraction of conformance constraints from the EU taxonomy. First, we outline our experimental setup in \autoref{sec:exp_setup}. Next, we apply our approach to the EU taxonomy and report the results in \autoref{sec:results}. Finally, we show the validity of our approach in \autoref{sec:validation}.

\subsection{Experimental Setup}
\label{sec:exp_setup}
We implemented the pipeline using Python 3 and Pandas in a Jupyter Notebook.%
The entire implementation, as well as all data we used and generated, is made available online for reproducibility purposes.\footnote{\url{https://github.com/fyndalf/unlocking-sustainability-compliance-replication-package}}
Additionally, the LLM we used in this study was Meta's Llama3,\footnote{\url{https://llama.meta.com/llama3/} [Accessed: 18/06/2024]} which is openly available. More precisely, we used \emph{llama-3-8b-instruct}, which is a version of Llama3 with 8 billion parameters that is fine-tuned for following instructions. Our prompting followed the official Llama3 documentation. For guaranteeing a quasi-deterministic output, we followed guidance on model-specific settings (such as \emph{temperature}, \emph{seed}).
Instead of self-hosting Llama3, we opted for using a web-based service (\url{https://groq.com}), which at the time of writing offered free API access to a hosted instance of Llama3. However, our implementation can easily be adapted to access other hosts or LLMs.

\subsection{Results}
\label{sec:results}
\paragraph{Constraint Types.} When analyzing the process constraints extracted with our pipeline, we observe the following distribution of constraint types: %
Out of a total of 1636 constraints we identified, we see a large focus on control-flow constraints (activity existence: 624; between: 8). For example, in order to make a substantial contribution to climate mitigation, market research and development business practices in the area of emission reduction technologies are required to have obtained a permit for operating a demonstration site, which can be interpreted as an activity existence constraint. Further, to contribute to the climate mitigation objective, business practices concerned with the construction, extension and operation of waste water collection and treatment must under some circumstances conduct an assessment of greenhouse gas emissions and subsequently disclose the result to investors (which can be understood as a control-flow constraint).
This is followed by process-irrelevant constraints (323). For example, the substantial contribution criteria to  climate mitigation of operating and providing personal mobility devices logistics requires that the vehicles are allowed to operate on the same infrastructure as bicycles and pedestrians, which is related rather to the environment in which the business practice is conducted, and not the business practice itself.
The third-most present constraint type is the one of data constraints, both within process activities (284) and between them (255). For example, the substantial contribution criteria for climate mitigation of the manufacturing of iron and steel describes greenhouse gas	emission threshold for individual steps of the manufacturing process. As another example, the DNSH to pollution criteria for the biodiversity objective of conservation and environmental protection requires the use of fertilizers across the entire business practice to be minimized, which can be understood as a data constraint between activities.
Less common are resource constraints (within: 96; between: 2) such as the substantial contribution criteria for the circular economy objective of the business practice of preparing end-of-life products and components for re-use, which makes requirements towards the tools and equipment being used. Finally, temporal constraints (within: 8; between: 36) are the least common: 
For example, the business practice of providing solutions for flood and drought risk prevention and protection is required to review the implemented solution periodically, in order to make a substantial contribution to the water protection objective.

\begin{figure}[htbp]
    \centering
    \vspace{-1em}
    \includegraphics[width=0.6\textwidth]{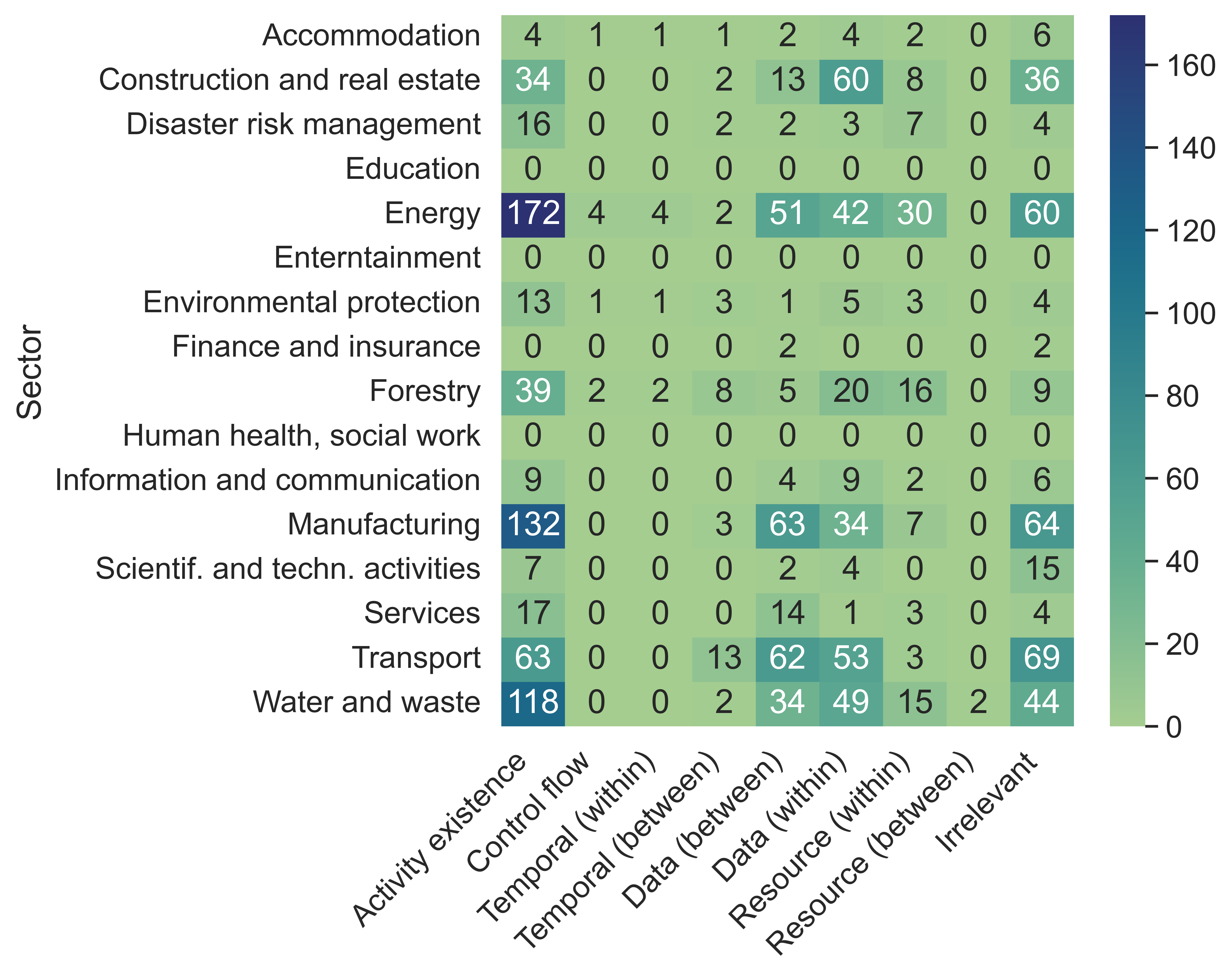}
    \caption{Constraint types per industry sector across all environmental objectives and screening criteria}
    \label{fig:constraint_type_per_sector}
    \vspace{-1em}
\end{figure}

\paragraph{Industry Sectors.} Looking at the individual industry sectors, we can further observe several patterns in the constraint types. Figure~\ref{fig:constraint_type_per_sector} depicts the sectors and constraints identified across all environmental objectives and screening criteria. First, we observe that energy, manufacturing, water, and transport are the industries most often constrained as to their contribution towards one or more climate goals. Noticeably, we see that activity existence criteria (such as permits that need to be obtained, assessments that need to be conducted) and data constraints (such as greenhouse gas limits, energy usage limits, etc.) play a vital role in assessing the taxonomy alignment in these sectors. Second, we see that resource constraints between activities, as well as control flow and temporal constraints play less of a role across all sectors. In the finance and insurance sector, we only identified two process-relevant constraints and two irrelevant ones. Finally, we see that three sectors, namely education, entertainment and human health/social work, have no identified constraints at all. A manual investigation into the taxonomy reveals that for both sectors, the taxonomy only provides substantial contribution criteria to the climate adaption goal, and no DNSH criteria. The substantial contribution criteria are provided in a very abstract manner, and we were unable to manually identify fine-grained process constraints. 

\begin{figure}[htbp]
    \centering
    \includegraphics[width=0.5\textwidth]{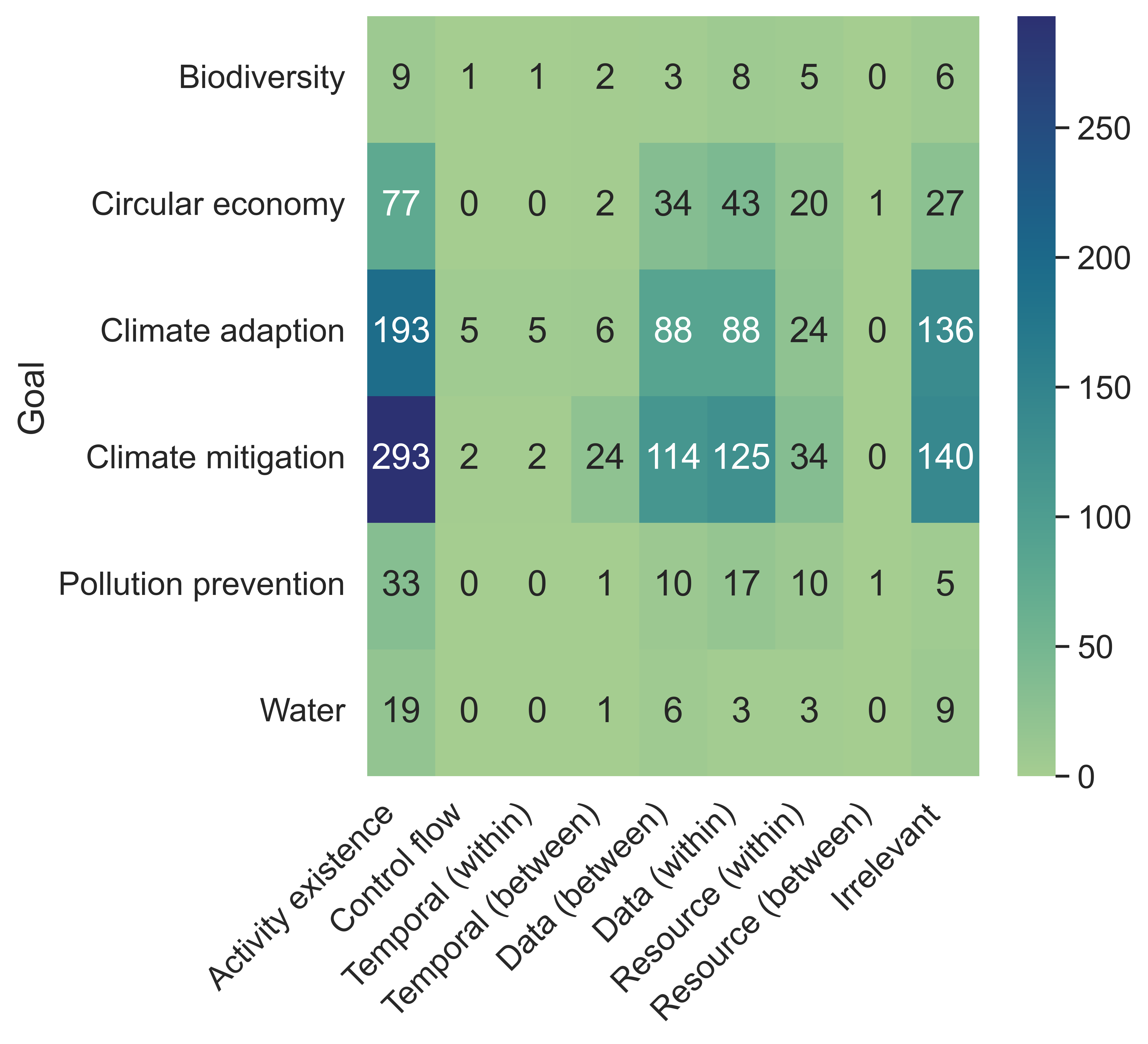}
    \caption{Number of constraints per type and environmental objective}
    \label{fig:constraint_type_per_goal}
    \vspace{-1em}
\end{figure}

\paragraph{Environmental Objectives.} Next, we analyze the constraint types belonging to the screening criteria of different environmental goals (meaning, the respective substantial contribution criteria of one goal and the associated DNSH criteria for all other goals). Figure \ref{fig:constraint_type_per_goal} illustrates the number of constraints of each type per environmental objective.
Here, we see that climate adaption and climate mitigation --- which are the two environmental objectives with the highest number of business practices for which they govern taxonomy alignment --- also contain the highest number of constraints across all types. Interestingly, we see that constraints related to the goals of pollution prevention and water protection are not at all characterized w.r.t. control-flow or temporal aspects within activities, while some constraints for the biodiversity, climate adaption, and climate mitigation use these process dimensions. Across all environmental objectives, however, we see a general focus on activity existence and data constraints, with resource constraints within activities also being represented. 

\subsection{Validation}
\label{sec:validation}

For validating our approach and the resulting insights, a research assistant with a background in environmental and resource management and one of the authors with a background in computer science and BPM manually assessed a sample of the taxonomy and constraint characterization (i.e., all 357 characterizations for the environmental objectives of water protection, circular economy, pollution prevention, biodiversity protection).
We compared each characterization and the underlying response of the LLM with the taxonomy's original text and assessed whether the result was \emph{entirely plausible} (we deem that all constraints have been found and classified accurately), \emph{largely plausible} (we deem that all constraints have been found, with a slight deviation in the constraint types; such as when a constraint that can be read as an activity existence constraint requiring an activity to be executed has instead been read as a resource constraint requiring the activity to be executed by a specific role or resource), \emph{somewhat plausible} (we deem that most relevant constraints have been found, but their classification is debatable), or \emph{implausible} (we deem that central constraints have not been found, or constraints have been clearly mischaracterized). Conflicts regarding the plausibility assessment were resolved in discussions.
This mode of validation allows us to assess the results of our approach without creating a ``gold standard'' result (as other studies do, see e.g.,~\cite{saiWhichLegalRequirements2024}), since we lack the regulatory expertise which would be necessary for creating such a standard.

Table~\ref{tab:evaluated_constraints} shows the share of constraints we assessed regarding their plausibility. We see that 340 of 357 characterizations are \emph{at least} assessed as somewhat plausible. Distributing the plausibility on a four-step scale (three being entirely plausible, zero being implausible), we see an \textbf{average plausibility of 2.74}, i.e., more than largely plausible. This means that, in general, we expect a characterization to be at least largely plausible. While we observed implausible characterizations, they seem to largely stem from references to taxonomy-external regulatory texts and standards which have not been considered and ambiguous terminologies (such as ``times'' as a frequency instead of referring to a temporal aspect). Overall, we infer that the classification approach can serve as a starting point for creating prescriptive models, and that it provides largely plausible constraint characterizations, which may need to be supplemented with a manual investigation. This is particularly the case when external regulations are involved.

\vspace{-1em}
\begin{table}[htbp]
    \centering
    \caption{Plausibility assessment of 357 constraint characterizations}
    \footnotesize
    \begin{tabular}{|l||c|c|c|c|}\hline
        Assessment & Entirely Plausible & Largely Plausible & Somewhat Plausible & Implausible\\\hline
        Characterizations & 308 & 24 & 8 & 17\\\hline
    \end{tabular}
    \normalsize
    \label{tab:evaluated_constraints}
    \vspace{-1em}
\end{table}

\section{Discussion}
\label{sec:discussion}

After presenting and validating our results, we now discuss them further.
As we have seen, business practices in the industry sectors of energy, manufacturing, transport, and water and waste constitute a large part of the process-relevant constraints we identified and thus appear well-suited to be investigated for their taxonomy alignment with conformance checking. In some sectors, particularly finance, education, entertainment, human health and social work, we were unable to identify a high number of constraints that would have been operationalizable for conformance checking. Therefore, we conclude, that these sectors appear less promising for applications of compliance monitoring with conformance checking.

In general, we can apply conformance checking to around 80\% (i.e., 1313
of 1636) of the constraints which we identified in the taxonomy. For all other constraints, and in particular, in sectors where we had difficulties automatically identifying constraints, conducting manual compliance monitoring and taking further expert knowledge into account appears indispensable. Hence, we have answered RQ2.

Further, as we have shown, the approach we have designed can aid in the creation of prescriptive models for compliance monitoring with conformance checking by helping end users to better understand: 1) what types of constraints are likely to be present in a single relevant piece of regulation (since it can be difficult to determine the concrete constraint type, or whether constraint is actually operationalizable), and 2) what types of constraints comprise a larger set of regulations (as it helps in choosing and implementing correct techniques). As a subsequent step, end users then need to operationalize the constraints into a prescriptive model for automatic compliance monitoring with conformance checking, drawing on existing approaches for this, in relation to actual business processes.
This allows the taxonomy to be operationalized for business processes. Therefore, we have addressed RQ1.

Taking a broader look at relevant BPM techniques, we believe that a particular focus on greenhouse gas emissions as data constraints, especially in the sectors of manufacturing, transport and energy, gives new importance to BPM approaches focussed on assessing emissions of business processes on process and activity levels (see, e.g.,~\cite{reckerMeasuringCarbonFootprint2011,reckerModelingAnalyzingCarbon2012,klessascheckSOPAFrameworkSustainabilityOriented2024}).
Broader still, the taxonomy itself has been the subject of various criticisms. On the one hand, the taxonomy's underlying notion of sustainable development \cite{alessiEUSustainabilityTaxonomy2019} has been criticized as ambiguous and an \textit{oxymoron}~\cite{johnstonReclaimingDefinitionSustainability2007}, and counterproductive to \emph{actual} sustainability~\cite{purvisThreePillarsSustainability2019}. On the other hand, the taxonomy has also been described as \emph{too} restrictive and as a bureaucratic burden that would be unable to benefit the overall economy~\cite{koothsEUTaxonomyMission2023}. Hence, the role played by the taxonomy in promoting sustainability is still subject to scholarly debate.

Nonetheless, this is one of the first papers to bring an understanding of the taxonomy to the business process management and enterprise computing communities. We have striven to provide conceptual clarity and impulses for future research on the taxonomy and its potential for sustainable business practices. Further, we believe that the pipeline we developed can potentially be applied to other regulatory frameworks as well.

\paragraph{Threats to Validity.}
There are several threats to the validity of our study.
First, we have not validated the constraint characterization in its entirety, and have rather focussed on plausibility instead of completeness. However, our experimental validation showed that the characterization is generally plausible, and can serve as a \emph{starting point} for further manual investigation.
Second, for compliance monitoring with conformance checking, the prescriptive process model into which regulations are operationalized needs to be shown to be regulatory compliant as well (see \cite{groefsemaUseConformanceCompliance2022}). This is a concern explicitly not addressed in our approach, as we have investigated in how far a prescriptive process model can be created at all. 
Moreover, technical limitations of LLMs, such as “confabulations”, “hallucinations” and inherent biases (see, e.g., ~\cite{suiConfabulationSurprisingValue2024,ladkinInvolvingLLMsLegal2023,benderDangersStochasticParrots2021}), apply to our study as well. However, our application is concerned with a very technical lens and is less of a generative application scenario. By following existing knowledge on prompt engineering in the BPM discipline, we sought to curb the impact of these limitations. Further, while our approach may produce results with some inaccuracies, the classification can still serve as a starting point for investigating individual constraints in depth.
Finally, other techniques in the compliance checking space exist that may allow further kinds of constraints to be extracted from regulations against which processes can be checked. However, as our investigation has been concerned only with conformance checking applications --- which are limited to the four process dimensions of control-flow, time, resources, and data --- we did not consider them further.

\section{Conclusion and Future Work}
\label{sec:conclusion-futur-work}

To conclude, in this paper we have investigated if the EU taxonomy for sustainable activities can be operationalized for automatic compliance monitoring. For this, we have developed a pipeline that uses few-shot learning with an LLM, to identify and characterize the types of constraints applicable for conformance checking. We saw that many constraints of various industries can, in fact, be operationalized for this, which will allow companies to automatically monitor compliance with regard to the taxonomy. We have demonstrated that operationalizing the EU taxonomy into constraints for conformance checking may be partially automated; in this paper, we provide a starting point for such an automation.
Besides this technical contribution, we have also introduced the taxonomy to the business process management and enterprise computing communities. The characterization pipeline may provide beneficial for assessing the capability of other complex regulatory frameworks as well.

Future work includes translating our classification approach into real-world application scenarios ---  further investigating how stakeholders can be supported in creating prescriptive models in line with the EU taxonomy, and how event log capturing and extraction benefits from our constraint classification would be a valuable contribution. We also aim to conduct more in-depth empirical evaluations of our mapping approach by comparing automatically generated results with results generated by taxonomy experts.
Moreover, since we focus exclusively on business process compliance and have abstracted away from constraints regarding their broader context, we have explicitly excluded the analysis of minimum safeguards. However, approaches that enable compliance monitoring of constraints across organizations exist (such as \cite{fdhilaDecomposition2020,knupleschDetecting2015,knupleschFramework2017}), and fruitful future work might lie in operationalizing the regulations and guidelines relevant to the taxonomy's minimum safeguard criterion for these approaches.
Finally, we believe that providing concrete guidance to end users in the form of a ``handbook'' or constraint patterns based on our preliminary findings reported herein would be a relevant addition.

\begin{credits}
\subsubsection{\ackname} This study was supported by the Deutsche Forschungsgemeinschaft (DFG, German Research Foundation) under grants no. 465904964, 496119880 and 531115272, by the German Federal Ministry of Education and Research (BMBF) under grant no. 16DII133 (Weizenbaum Institute), and by the Einstein Foundation Berlin under grant no. EPP-2019-524. Further, the authors would like to thank Man Tuen Chan for his support in validating the approach presented in this paper.

\subsubsection{\discintname} The authors have no competing interests to declare that are relevant to the content of this article.
\end{credits}

\bibliographystyle{splncs04}
\bibliography{main}
\end{document}